Paper 08-0056

# Analysis of the Effect of Speed Limit Increases on Accident-Injury Severities

by

Nataliya V. Malyshkina
Research Assistant, School of Civil Engineering
550 Stadium Mall Drive
Purdue University
West Lafayette, IN 47907-2051
nmalyshk@purdue.edu

Fred Mannering
Professor, School of Civil Engineering
550 Stadium Mall Drive
Purdue University
West Lafayette, IN 47907-2051
flm@purdue.edu

**Abstract**

The influence of speed limits on roadway safety has been a subject of continuous debate in the State of Indiana and nationwide. In Indiana, highway-related accidents result in about 900 fatalities and forty thousand injuries annually and place an incredible social and economic burden on the state. Still, speed limits posted on highways and other roads are routinely exceeded as individual drivers try to balance safety, mobility (speed), and the risks and penalties associated with law enforcement efforts. The speed-limit/safety issue has been a matter of considerable concern in Indiana since the state raised its speed limits on rural interstates and selected multilane highways on July 1, 2005. In this paper, the influence of the posted speed limit on the severity of vehicle accidents is studied using Indiana accident data from 2004 (the year before speed limits were raised) and 2006 (the year after speed limits were raised on rural interstates and some multi-lane non-interstate routes). Statistical models of the injury severity of different types of accidents on various roadway classes were estimated. The results of the model estimations showed that, for the speed limit ranges currently used, speed limits did not have a statistically significant effect on the severity of accidents on interstate highways. However, for some non-interstate highways, higher speed limits were found to be associated with higher accident severities – suggesting that future speed limit changes, on non-interstate highways in particular, need to be carefully assessed on a case-by-case basis.

**Introduction**

The speed-limit changes that took effect on July 1, 2005, made Indiana the 30th U.S. state to raise interstate speed limits up to 70 mi/h (raising them from 65 mi/h on rural interstates). Speed limits were also increased on some multilane non-interstate highways. This intensified a statewide debate on the tradeoff between highway mobility (speed) and safety. This debate has raged throughout the US for more than three decades since the passage of the Emergency Highway Energy Conservation Act in 1974, which mandated the 55 mi/h national maximum speed limit on interstate highways in the US. State and federal speed-limit policy changes have been fueled by various research findings and subsequent legislation, such as the National Highway System Designation Act of 1995 that gave states complete freedom to set interstate speed limits.

With regard to safety, most research efforts have concluded that the 1974-mandated 55 mi/h interstate speed limit had saved lives (*1*, *2*). This has been confirmed by some studies that have looked at recent speed limit increases on interstates. As an example, Kockelman and Bottom (*3*) found that a speed limit increase from 55 to 65 mi/h resulted in roughly a 3% increase in the accident rate and a 24% increase in the probability of a fatality once an accident occurred. For speed limit increases from 65 to 75 mi/h, they found a 0.64% increase in the accident rate and in a lower 12% increase in the probability of fatal injury once an accident occurred. The authors speculated that these lower percentage increases from 65 to 75 mi/h speed limits (relative to the 55 to 65 mi/h speed-limit increases) may have been the result of drivers' heightened awareness of risk at higher speeds or that roads assigned the higher 75 mi/h in their study's sample may have been inherently safer.

However, other studies have contended that legislation-enabled speed-limit increases have actually saved lives. As an example, Lave and Elias (*4*) argued that the increase from 55 mi/h to 65 mi/h saved lives because of shifts in law enforcement resources, the ability of higher-speed-limit interstates to attract riskier drivers away from inherently more dangerous non-interstate highways, and possible reductions in speed variances.

Understanding the magnitude of the safety impacts of increasing speed limits, or even the direction of safety impacts (whether safety is improved or compromised), remains a contentious subject because research has not been able to convincingly unravel the impacts of speed limit changes from the confounding effects of time-varying changes in factors such as highway enforcement, vehicle miles traveled, vehicle occupancy, seat belt usage, alcohol use and driving, vehicle fleet mix (proportions of passenger cars, minivans, pickup trucks, and sport utility vehicles), vehicle safety features (increasing adoption of air bags, antilock brakes, other active safety systems), speed limits on other road classes and in other states, driver expectations, and driver adjustment and adaptation to risk.

In this paper, the recent increase in Indiana speed limits was studied by undertaking a statistical analysis to assess the effect that speed limits have on roadway safety by considering the relationship between speed limits and observed accident-injury severities. In assessing the impact of speed limits on accident-injury severities, the injury level sustained by the most critically injured individual in an accident was used to conduct appropriate statistical tests to determine whether the posted speed limit had any significant effect on these injury severities and whether the possible effect changed after the speed limits were raised.

**Methodology**

Accident severity (the most severe injury sustained by any vehicle occupant in the accident) has discrete outcomes ranging from property damage only, to injury, to fatality. Given that these severity data are ordered responses from less severe to more severe, an ordered probability model would seem to be a natural approach and one that has been successfully applied to accident severity analysis by many researchers (*5, 6, 7, 8, 9, 10, 11, 12*). However, an alternative to an ordered model is the unordered probability approach that includes multinomial, nested and mixed logit models (see *13, 14, 15, 16, 17, 18, 19, 20, 21*). Relative to ordered probability models, traditional multinomial, nested and mixed logit structures do not account for the ordering of injury-severity data. However, multinomial, nested and mixed logit models do offer a more flexible functional form. For example, the traditional multinomial logit can provide consistent parameter estimates in the presence of the possible underreporting of accidents (for example, a known problem with police-reported data is the underreporting of property damage only accidents). Also, they can relax the parameter restriction imposed by ordered probability models that does not allow a variable to simultaneously increase (or decrease) both high and low injury severities. This monotonic effect of variables imposed by ordered probability models and its potential adverse consequences is discussed in Eluru and Bhat (*22*), Bhat and Pulugurta (*23*), and Washington et. al. (*24*). In this paper a multinomial logit model for accident injury-severity outcomes was used.

For the multinomial logit formulation, a linear function of covariates that determine the probability of the accident injury outcome (of the most severely injured person) being reported as property damage only (no injury), injury or fatality is defined as:

$$Z_{in} = \boldsymbol{\beta}_i \boldsymbol{X}_n + \varepsilon_{in}, \tag{1}$$

where $Z_{in}$ is a linear function determining the probability of accident severity outcome $i$ for accident $n$, $\boldsymbol{X}_n$ is a vector of measurable characteristics for accident $n$ that determine outcome $i$ (such as speed limit, driver characteristics, roadway characteristics, environmental conditions, etc.), $\boldsymbol{\beta}_i$ is a vector of estimable parameters, and $\varepsilon_{in}$ is an error term that accounts for unobserved factors influencing resulting outcomes. McFadden (*25*) has shown that if $\varepsilon_{in}$ are assumed to be generalized extreme value distributed, the standard multinomial logit model results,

$$P_n(i) = \frac{\exp[\boldsymbol{\beta}_i \boldsymbol{X}_n]}{\sum_I \exp[\boldsymbol{\beta}_I \boldsymbol{X}_n]}, \tag{2}$$

where $P_n(i)$ is the probability that accident $n$ has accident-severity outcome $i$ and $I$ is the set of possible outcomes. This model is estimable by standard maximum likelihood methods (*24*).

To assess the effect of the vector of estimated parameters ($\boldsymbol{\beta}_i$), elasticities are computed for each accident $n$ ($n$ subscripting omitted) as

$$E_{x_{ki}}^{P(i)} = \frac{\partial P(i)}{\partial x_{ki}} \times \frac{x_{ki}}{P(i)}, \quad (3)$$

where $P(i)$ is the probability of discrete outcome $i$ and $x_{ki}$ is the value of variable $k$ for outcome $i$. This gives (using Equation 2 and 3),

$$E_{x_{ki}}^{P(i)} = \left[1 - P(i)\right] \beta_{ki} x_{ki}, \quad (4)$$

where $\beta_{ki}$ is the estimated parameter associated with variable $x_{ki}$. Elasticity values can be roughly interpreted as the percent effect that a 1% change in $x_{ki}$ has on the accident-severity outcome probability $P(i)$.

The statistical analysis also used multiple statistical tests to determine if the accident data should be split by roadway type (rural interstate, urban interstate, rural arterials, etc.) and accident type (single vehicle, car colliding with car, car colliding with truck, etc). For this analysis a likelihood ratio test was used. The appropriate test statistic is (*24*),

$$-2\left[LL(\boldsymbol{\beta}) - \sum_{m=1}^{M} LL(\boldsymbol{\beta}_m)\right] \sim \chi^2_{df = (M-1) \times K}, \quad (5)$$

where $LL(\boldsymbol{\beta})$ is the log-likelihood at converged values of $\boldsymbol{\beta}$ for the model estimated for the whole data sample; $LL(\boldsymbol{\beta}_m)$ is the log-likelihood of the model estimated for observations in the $m^{th}$ data subset (road/accident type combinations) and $\boldsymbol{\beta}_m$ is the vector of parameters estimated for this model ($m = 1, 2, 3, ..., M$); $M$ is the number of the data subsets; $K$ is the number of parameters estimated for each model (i.e. $K$ is the number of estimated parameters in vectors $\boldsymbol{\beta}$ and $\boldsymbol{\beta}_m$); and there are $(M-1) \times K$ degrees of freedom. The null hypothesis for Equation 5 is that the $\boldsymbol{\beta}$'s in the $M$ subset models are the same. If this hypothesis can be rejected with high confidence, estimation of separate models for the data subsets is warranted.

**Data**

The accident data used in this study were from the Indiana Electronic Vehicle Crash Record System (EVCRS) which was launched in 2004 and includes available information on all accidents investigated by Indiana police. The information on accidents included into the EVCRS can be divided into three major categories: roadway and environmental data (including weather, roadway and traffic conditions, roadway geometrics, posted speed limits, etc.); vehicle data (including information on all vehicles involved in an accident, type and model of each vehicle, vehicle model year, etc.); and occupant data (including information such as age and gender on all people who were involved in the accident and their injury status). These data gave 127 variables for each accident. Data were considered from 204,382 accidents that were reported in the Indiana State accident databases for 2004 (the year before the speed limits were raised) and from 182,922 accidents that were reported in the Indiana State accident databases for 2006 (the year after the speed limits were raised).

Of the 182,922 accidents in the 2006 database, 65% occurred on locally maintained streets and roads (city and county), 16.7% on state routes, 10.9% on US routes, and 7.4% on urban and rural interstates. Of these, 52.9% were two-vehicle accidents involving only passenger vehicles (passenger car and light trucks which include sport utility vehicles, vans and pickup trucks), 3.9% were vehicle accidents involving a large truck and a passenger vehicle, 31.1% were single vehicle accidents, and 12.1% were other accident types (involving 3 or more vehicles). This compares closely to the 2004 database that showed 69.3% occurred on locally maintained streets and roads (city and county), 14.8% on state routes, 9.7% on US routes, and 6.2% on urban and rural interstates. Of these 2004 accidents, 54.7% were two-vehicle accidents involving only passenger vehicles (passenger car and light trucks which include sport utility vehicles, vans and pickup trucks), 4.8% were vehicle accidents involving a large truck and a passenger vehicle, 28.6% were single vehicle accidents, and 11.9% were other accident types (involving 3 or more vehicles)

Of the accidents in 2006, unsafe speed was identified as the primary cause in 5.78% of the accidents. This compares to 7.28% of the 2004 accidents that listed unsafe speed as the primary cause of the accident (before the increase in speed limits). In terms of injury severity levels in 2006, 79.03% were property damage only, 20.56% were injury and 0.41% were fatality (the numbers for 2004 were 78.53% were property damage only, 21.06% were injury and 0.41% were fatality).

With regard to the effect of speed limits, in 2006 unsafe speed was listed as the primary cause of the accident in 11.4% of the accidents on roads with 65mi/h and 70mi/h speed limits, 7.7% on roads with speed limits of 55mi/h and 60mi/h, 6.6% on roads with speed limits from 35mi/h to 50 mi/h, and 4.6% on roads with speed limits of 30 mi/h or less. This compares with 2004 data that showed that unsafe speed was listed as the primary cause of the accident in 19.4% of the accidents on roads with 65 mi/h speed limits (the maximum speed in 2004), 10.6% on roads with speed limits of 55 mi/h and 60 mi/h, 7.5% on roads with speed limits from 35mi/h to 50 mi/h, and 6.0% on roads with speed limits of 30 mi/h or less. The corresponding accident severity levels, by speed-limit category, for 2004 and 2006 are presented in Table 1. This table shows some variation but generally similar numbers between the years. However, to properly unravel this relationship, a multivariate analysis is needed to control for all of the many factors that can potentially affect this relationship.

**Estimation results**

Multinomial logit estimation results using the 2006 data are presented in Table 2 for the accident injury-severity models (the most severely injured occupant). The results for the 2004 data are similar and are not presented here to save space (please see Malyshkina et. al (*26*) for a complete presentation of the 2004 results). Again, the possible outcomes are property damage only, injury and fatality. For estimation purposes, the function determining property damage only (given by Equation 1) is set to zero without loss of generality (*24*). Based on the results of likelihood ratio tests, 34 different injury-severity models were estimated based on combinations of roadway type, roadway location and the number and types of vehicles involved in the accident. In addition to the speed-limit parameter estimates shown in Table 2, the models included a wide variety of variables that were found to significantly influence injury severity including seasonal indicators, day-of-week indicators, peak-hour indicators, other time of day indicators, construction zone indicators, lighting conditions, precipitation indictors (snow, rain, clear, etc.), pavement condition indictors (dry, wet, ice, etc.), median type, presence of vertical and horizontal curves, vehicle age, number of vehicle occupants, traffic control indicators, driver age, driver gender.

Turning specifically to the speed-limit parameter estimates shown in Table 2, it is found that speed limits did not significantly affect accident-injury severities on interstate highways (this was also true of the model estimations based on 2004 data). While the reasons for this are not know for certain, there could be a number of contributing factors to this finding. One is the issue of speed limit compliance on interstate highways. In a survey of Indiana drivers conducted in the Fall of 2005 (a few months after Indiana interstate speed limits were raised), it was found that under free flow conditions drivers reported driving an average of nearly 11 mi/h over a 55 mi/h interstate speed limit, about 9 mi/h over a 65 mi/h interstate speed limit and less than 8 mi/h over a 70 mi/h speed limit (*27*). Thus it appears that driver behavior is compressing the effect of speed limits (a 15 mi/h increase in speed limits results in a less than 15 mi/h increase in speeds). Also, this same survey found that the standard deviation of self-reported free-flow speeds declined from roughly 6 mi/h on interstates posted 55 mi/h to about 5 mi/h on interstates posted 65 mi/h or 70 mi/h. The reduced standard deviation of speed may be mitigating the effect of the higher overall speeds on severity. There could also be behavioral elements involved such as drivers becoming more alert (perhaps with lower reaction times) at higher speeds and enabling them to take actions to reduce accident severity. Finally, the high design standards of the interstate system may be mitigating the effects of higher speeds. All of the above factors seem to be sufficient to offset the physics involved with higher travel speed. Whether this would be true for speed limits above 70 mi/h is an open question that our data can not support.

In contrast to interstates, Table 2 shows that for many other highway types, increases in speed limits significantly increase the likelihood that the accident will result in an injury or fatality (this was also true of the model estimations based on 2004 data). And, the elasticities show that the effect can be reasonably large. For example, as shown in Table 2, for rural-county-road accidents involving a car or light truck with a heavy truck, a 1% increase in the speed limits results in a 2.77% increase in the probability of fatality and a 2.35% increase in the probability of injury. For rural-state-route accidents involving a car or light truck with another car or light truck, a 1% increase in the speed limit results in a 11.9% increase in the probability of fatality and a 1.32% increase in the probability of injury. The accident-injury severity findings on non-interstate highways suggest that extreme caution needs to be exercised when raising the speed limits on these roads. It is speculated that the lack of access control, lower design standards and the greater demand placed on driver attention can make these highways quite sensitive to speed-limit changes.

**Temporal Stability: Interstates**

The fact that speed limits on interstates were not found to affect the severity of accidents is worth a closer look. The previous analysis is based on cross-sectional data comparing the effect of different speed limits on a class of roadways for a single year (2004 or 2006). This is an important consideration because it controls for possible changes in vehicle safety features, enforcement levels and driver behavior that may vary from one year to the next. Still, it may also be of interest to compare 2004 and 2006 interstate accident-severity models to see if the 2005 speed-limit increase (and the other factors that may have varied over this time period) significantly changed the estimated parameters in the accident severity models. To address this possibility, the following are considered: 1) interstates that had 55 mi/h speed limits in 2004 and remained at 55 mi/h in 2006 (which were interstates in urban areas); and 2) interstates that were 65 mi/h in 2004 and increased to 70 mi/h in 2006 (which were those in rural areas). Again, applying the likelihood ratio test, the appropriate statistical test is (*24*),

$$-2\left[LL(\boldsymbol{\beta}_{all}) - LL(\boldsymbol{\beta}_{2004}) - LL(\boldsymbol{\beta}_{2006})\right] \sim \chi^2_{df\,=\,(K_{2004}+K_{2006})-K_{all}}, \qquad (6)$$

where $LL(\boldsymbol{\beta}_{all})$ is the log-likelihood at converged values of $\boldsymbol{\beta}$ for the model estimated 2004 and 2006 data for the accident type being considered; $LL(\boldsymbol{\beta}_{2004})$ is the log-likelihood of the model estimated for 2004 observations; $LL(\boldsymbol{\beta}_{2006})$ is the log-likelihood of the model estimated for 2006 observations; $K_{all}$ is the number of parameters in $\boldsymbol{\beta}_{all}$; $K_{2004}$ is the number of parameters in $\boldsymbol{\beta}_{2004}$; and $K_{2006}$ is the number of parameters in $\boldsymbol{\beta}_{2006}$. The null hypothesis for Equation 6 is that the $\boldsymbol{\beta}$'s in the 2004 and 2006 are the same. If this hypothesis can be rejected, it would indicate that the parameters have shifted from 2004 to 2006, suggesting that the increased speed limit and possibly other influences (such as changes in driver behavior, changes in enforcement levels, and changes in vehicle safety features) may have changed the effect of factors that determine injury severity.

For interstates that had the same 55 mi/h speed limits in 2004 and 2006, the application of this likelihood ratio test to the various models indicated that the accident severity parameter estimates (for single and multivehicle crashes) did not significantly change from 2004 to 2006. In no case could the null hypothesis be rejected even at a very modest 70 percent confidence level. This indicates that the "halo effect" of the increased speed limits (from 65 mi/h to 70 mi/h in rural areas) did not significantly affect those urban interstates that remained at 55 mi/h.

For interstates that were 65 mi/h in 2004 and increased to 70 mi/h in 2006, temporal stability tests again showed that accident severity parameter estimates (for single and multivehicle crashes) did not significantly change from 2004 to 2006 (the null hypothesis could not be rejected in any of the cases even at the very modest 70 percent confidence level). The temporal stability of these interstate models adds further evidence to support the cross-sectional finding that the higher range of speed limits in effect on Indiana interstates in 2006 has not significantly affected the severity of accidents.

**Summary and Conclusions**

The findings of this study are drawn from multinomial models of accident severity defined by the injury level of the most severely injured person in the accident. The estimation results found that speed limits did not significantly affect accident injury severities on interstate highways. This is an important finding because the July 1, 2005 increase in maximum interstate speeds from 65 mi/h to 70 mi/h has been the focus of considerable media attention. One can speculate that this finding is a result of a number of factors including possible reductions in speed variance as speed limits increase, driver responses to higher speed limits and the high design standards of the interstate system which appear to be able to accommodate modest increases in speed limits.

For non-interstate highways, the results are quite different. For non-interstate highways, the accident data show that higher speed limits are associated with a greater likelihood of injury and/or fatality on some (but not all) roadway types (county, state, city and US routes) and accident types (single and two-vehicle).

This study's findings have a number of implications for speed limit polices in the State of Indiana. With regard to interstate speeds, the findings suggest that the effect of speed limits on accident injury severity are not necessarily a cause for concern for the speed-limit ranges that were considered in this study (55 mi/h to 70 mi/h). Whether this finding would hold true if speed limits were increased further to 75 mi/h or 80 mi/h (both of which would exceed the 70 mi/h interstate-standard design speed) remains an open question. To be sure, the additional speed would increase stopping distances and the energy that would need to be dissipated in the accident. Also, at some point, higher speed limits may start increasing the variance in driver speeds as some drivers continue to drive at or above the speed limit while others drive below the speed limit because the speed limit may have been raised above their "optimum" speed. With these factors considered (along with other factors that may come into play such as variations in driver behavior in response to speed limits), there is likely a point beyond which higher speed limits will significantly increase the severity of accidents on interstates.

With regard to speed limit policies on roadways other than interstate highways, our results suggest that considerable caution should be exercised. The findings show that, on some non-interstate roadway and accident type combinations, higher speed limits significantly increase the likelihood that accidents will result in injuries and fatalities. Thus, changing speed limits on non-interstate highways should be done on a case-by-case basis taking into account past accident history as well as the specific geometrics and access control of the facility, as these factors can vary widely even within the same class of highways (non-interstate).

## Acknowledgments

The authors were supported by the Indiana Department of Transportation/Joint Transportation Research Program (JTRP) Project – SPR 3030. The contents of this paper reflect the views of the authors, who are responsible for the facts and the accuracy of the data presented herein. The contents do not necessary reflect the official views or policies of the Federal Highway Administration and the Indiana Department of Transportation, nor do the contents constitute a standard, specification, or regulation. The comments and suggestions of Brad Steckler are gratefully acknowledged.

**List of Tables**

Table 1. Indiana accident injury-severity distributions by posted speed limits in 2004 and 2006.

Table 2. Speed limit parameter and elasticity estimates (computed for statistically significant parameters) for accident severity models based on 2006 Indiana accident data.

*Malyshkina and Mannering* 13Table 1.  Indiana accident injury-severity distributions by posted speed limits in 2004 and 2006.

|  | **Injury Severity Level** | | |
|---|---|---|---|
| **Speed limit** | **Property Damage Only** | **Injury** | **Fatality** |
| 2004 posted 65 mi/h | 81.7% | 17.7% | 0.6% |
| 2006 posted 65 mi/h or 70 mi/h | 81.9% | 17.4% | 0.7% |
| 2004 posted 55 mi/h and 60 mi/h | 76.7% | 22.3% | 1.1% |
| 2006 posted 55 mi/h and 60 mi/h | 77.8% | 21.3% | 0.9% |
| 2004 posted 35 mi/h to 50 mi/h | 74.5% | 25.2% | 0.4% |
| 2006 posted 35 mi/h to 50 mi/h | 75.3% | 24.3% | 0.4% |
| 2004 posted 30 mi/h or less | 80.6% | 19.2% | 0.2% |
| 2006 posted 30 mi/h or less | 82.1% | 17.8% | 0.2% |

Table 2. Speed limit parameter and elasticity estimates (computed for statistically significant parameters) for accident severity models based on 2006 Indiana accident data.

| Model (C=cars, LT=Light trucks; HT=Heavy Trucks) | | | Speed limit parameter estimate (*t*-ratio) | | Fatality Elasticity | Injury Elasticity |
|---|---|---|---|---|---|---|
| | | | fatality | injury | | |
| County road | Rural | (C/LT)+(C/LT) | 0.0396(5.48) | 0.0396(5.48) | 1.61 | 1.20 |
| | | (C/LT)+(HT) | 0.0648(3.06) | 0.0648(3.06) | 2.77 | 2.35 |
| | | one vehicle | 0.00506(2.04) | 0.00506(2.04) | 0.24 | 0.19 |
| | Urban | (C)+(C) | 0.00689(0.00) | 0.00507(.321) | | |
| | | (C)+(LT) | 0.0231(0.00) | 0.0613(2.43) | | 1.80 |
| | | (LT)+(LT) | .0110(0.00) | 0.0269(0.84) | | |
| | | (C/LT)+(HT) | -0.5454(-.518) | -0.0288(-0.30) | | |
| | | one vehicle | -0.0852(-1.23) | 0.000725(0.07) | | |
| Interstate | Rural | (C/LT)+(C/LT) | 0.103(1.28) | 0.00872(0.88) | | |
| | | (C/LT)+(HT) | 0.150(0.91) | 0.00133(0.06) | | |
| | | one vehicle | -0.0237(-1.55) | -0.0237(-1.55) | | |
| | Urban | (C/LT)+(C/LT) | 11.04(0.00) | -0.00108(-0.14) | | |
| | | (C/LT)+(HT) | -0.00188(-0.01) | 0.0120(0.52) | | |
| | | one vehicle | 0.00776(0.20) | 0.00384(0.48) | | |
| State route | Rural | (C/LT)+(C/LT) | 0.248(3.48) | 0.0416(3.25) | 11.9 | 1.32 |
| | | (C/LT)+(HT) | 0.127(2.50) | 0.127(2.50) | 5.79 | 5.36 |
| | | one vehicle | 0.0636(2.34) | 0.0127(2.25) | 3.34 | |
| | Urban | (C/LT)+(C/LT) | 0.251(3.35) | .0290(7.95) | 9.40 | 0.84 |
| | | (C/LT)+(HT) | 5.60(0.00) | 0.452(1.73) | | |
| | | one vehicle | 0.0268(0.42) | -0.0115(-0.83) | | |
| City street | Rural | (C/LT)+(C/LT) | 0.0414(6.13) | 0.0414(6.13) | 1.46 | 1.12 |
| | | (C/LT)+(HT) | 0.0185(0.00) | 0.540(1.43) | | |
| | | one vehicle | -0.0800(-1.46) | -0.00409(-0.38) | | |
| | Urban | (C)+(C) | 0.0251(6.33) | 0.0251(6.33) | 0.81 | 0.63 |
| | | (C)+(LT) | 0.0218(4.65) | 0.0218(4.65) | 0.73 | 0.56 |
| | | (LT)+(LT) | 0.0343(4.20) | 0.0343(4.20) | 1.14 | 0.87 |
| | | (C/LT)+(HT) | 0.0284(2.34) | 0.0284(2.34) | 0.94 | 0.83 |
| | | one vehicle | 0.00968(0.38) | -0.00128(-0.24) | | |
| US route | Rural | (C/LT)+(C/LT) | 0.0644(1.32) | 0.0272(1.84) | | |
| | | (C/LT)+(HT) | 0.0608(3.07) | 0.0608(3.07) | 3.12 | 2.28 |
| | | one vehicle | 0.0137(1.56) | 0.0137(1.56) | | |
| | Urban | (C/LT)+(C/LT) | 0.0154(2.14) | 0.0154(2.14) | 0.61 | 0.44 |
| | | (C/LT)+(HT) | 0.0586(3.60) | 0.0586(3.60) | 2.33 | 2.02 |
| | | one vehicle | 0.0327(0.45) | 0.0134(.878) | | |